\documentclass[11pt,onecolumn]{article}
\usepackage[utf8]{inputenc}
\usepackage{amsmath,amsfonts,amssymb,graphicx,authblk,placeins,cite}
\usepackage{comment}
\usepackage{color,soul}
\usepackage{hyphenat}
\usepackage[top=50pt,bottom=50pt,left=70pt,right=70pt]{geometry}
\usepackage{sectsty}
\usepackage[font=sf]{caption}
\usepackage{hyperref}

\begin{document}
\allsectionsfont{\sffamily}
\pagenumbering{gobble}

\title{\sffamily \Huge Neural Network based Formation of Cognitive Maps of Semantic Spaces and the Emergence of Abstract Concepts
 \vspace{1 cm}}


\author[1,2]{\sffamily Paul Stoewer}
\author[1,3]{\sffamily Achim Schilling}
\author[2]{\sffamily Andreas Maier}
\author[1,2,3,4,*]{\sffamily Patrick Krauss}


\affil[1]{\small{Cognitive Computational Neuroscience Group, University Erlangen-Nuremberg, Germany}}

\affil[2]{\small{Pattern Recognition Lab, University Erlangen-Nuremberg, Germany}}

\affil[3]{\small{Neuroscience Lab, University Hospital Erlangen, Germany}}

\affil[4]{\small{Linguistics Lab, University Erlangen-Nuremberg, Germany}}

\affil[*]{\small{corresponding author}}

\maketitle


\begin{abstract}{\sffamily \noindent
How do we make sense of the input from our sensory organs, and put the perceived information into context of our past experiences? The hippocampal-entorhinal complex plays a major role in the organization of memory and thought. The formation of and navigation in cognitive maps of arbitrary mental spaces via place and grid cells can serve as a representation of memories and experiences and their relations to each other. The multi-scale successor representation is proposed to be the mathematical principle underlying place and grid cell computations. Here, we present a neural network, which learns a cognitive map of a semantic space based on 32 different animal species encoded as feature vectors. The neural network successfully learns the similarities between different animal species, and constructs a cognitive map of 'animal space' based on the principle of successor representations with an accuracy of around 30\% which is near to the theoretical maximum regarding the fact that all animal species have more than one possible successor, i.e. nearest neighbor in feature space. Furthermore, a hierarchical structure, i.e. different scales of cognitive maps, can be modeled based on multi-scale successor representations. We find that, in fine-grained cognitive maps, the animal vectors are evenly distributed in feature space. In contrast, in coarse-grained maps, animal vectors are highly clustered according to their biological class, i.e. amphibians, mammals and insects. This could be a possible mechanism explaining the emergence of new abstract semantic concepts. Finally, even completely new or incomplete input can be represented by interpolation of the representations from the cognitive map with remarkable high accuracy of up to 95\%. We conclude that the successor representation can serve as a weighted pointer to past memories and experiences, and may therefore be a crucial building block to include prior knowledge, and to derive context knowledge from novel input. Thus, our model provides a new tool to complement contemporary deep learning approaches on the road towards artificial general intelligence. 
}
\end{abstract}

\newpage
\section*{Introduction}
\begin{figure}[htbp]
    \centering
    \includegraphics[width = 13cm]{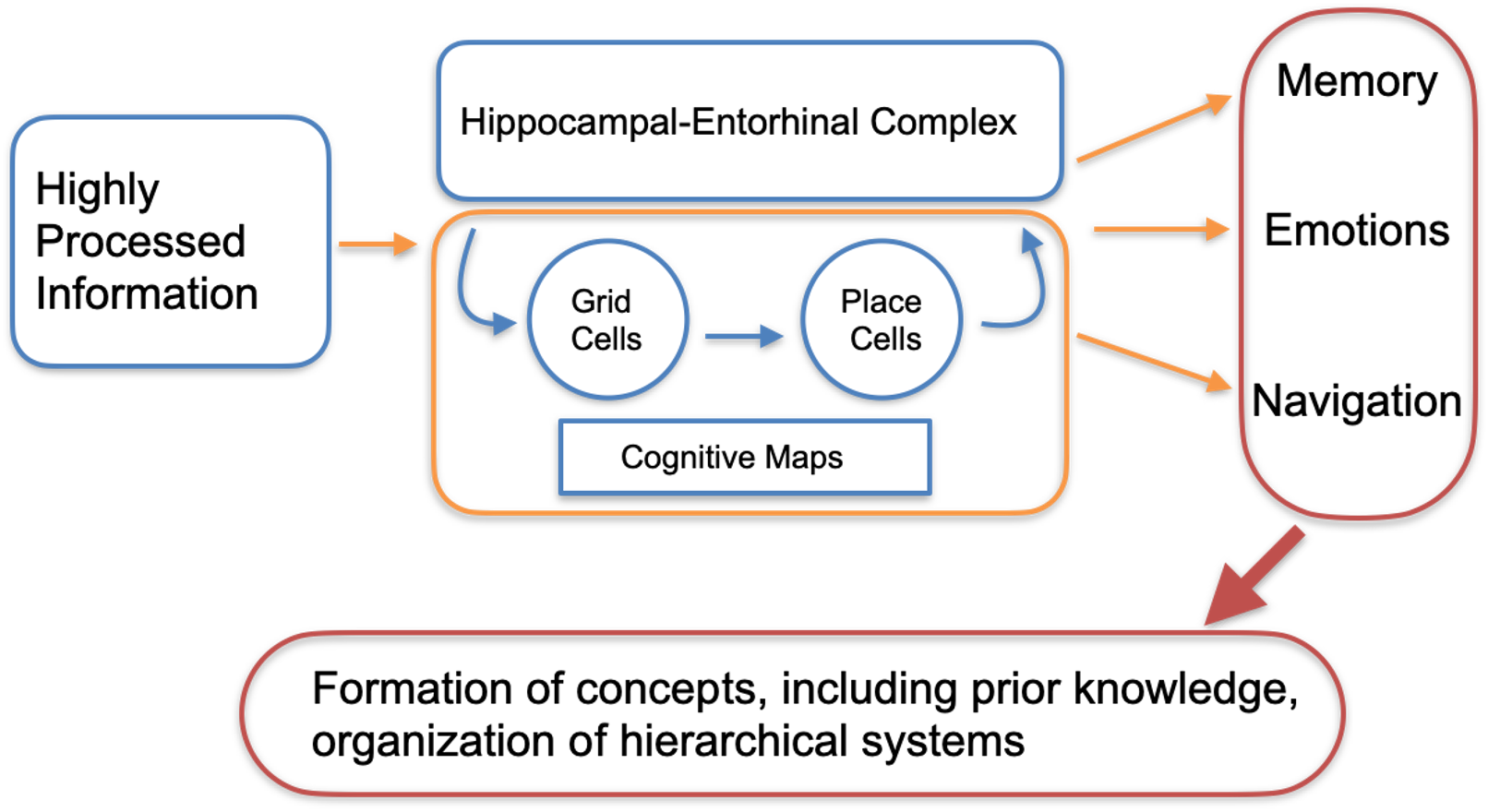}
    \caption{\textbf{Simplified sketch model of the hippocampal-entorhinal complex.} Highly processed information is fed into the system and becomes associated with existing memories and past experiences. Place and grid cells enable the formation of map like codes, and finally cognitive maps. The complex also supports navigation, emotions, the formation of concepts, inclusion of prior knowledge, and the organization of hierarchical representations.}
    \label{fig1}
\end{figure}

The hippocampal-entorhinal complex supports spatial navigation and forms cognitive maps of the environment \cite{okeefe_hippocampus_1971}. However, recent research suggests that formation of and navigation on cognitive maps are not limited to physical space, but extend to more abstract conceptual, visual or even social spaces  \cite{epstein2017cognitive,park2021inferences, killian_grid_2018}. A simplified processing framework for the complex can be described as following: highly processed information from our sensory organs are fed into the hippocampal complex where the perceived information is put into context, i.e. associated with past experiences \cite{opitz_memory_2014}. Grid \cite{hafting2005microstructure} and place \cite{o1971hippocampus} cells enable map like codes, and research suggests that they form cognitive maps \cite{o1978hippocampus}\cite{moser2017spatial}, thereby contributing to process memories, emotions and navigation \cite{kandel_principles_2013}(cf. \ref{fig1}).

Furthermore, it is known that the hippocampus plays a crucial role for episodic and declarative memory \cite{tulving1998episodic,reddy2021human}. However, whether memories are directly stored in the hippocampus, and how they are retrieved through the hippocampus, is depending on different theories still under discussion. Therefore, the exact role of the hippocampus in the domain of memory is still not fully understood \cite{kryukov_role_2008}. According to the \emph{multiple trace theory} \cite{nadel_memory_1997}, memories are not directly stored in the hippocampus. Instead, memory content is stored in the cerebral cortex, and the hippocampus forms representations of memory traces which can serve as pointers to retrieve memory content from the cerebral cortex.

Furthermore memory can be represented at different scales along the hippocampal longitude axis, like e.g. varying spatial resolutions \cite{collin2015memory}. In the context of spatial navigation the different scales serve to navigate with different horizons \cite{brunec2019predictive}. In the context of abstract conceptual spaces, different scales might correspond to different degrees of abstraction \cite{milivojevic2013mnemonic}. In general, multi-scale cognitive maps enable flexible planning, generalization and detailed representation of information \cite{momennejad2020learning}. 

Various different computational models try to explain the versatility of the hippocampal-entorhinal complex. One of these candidate models successfully reproduces the firing patterns of place and grid cells in a large number of different experimental scenarios, indicating that the hippocampus works like a predictive map based on multi-scale successor representations (SR) \cite{stachenfeld2014design, stachenfeld2017hippocampus,mcnamee2021flexible}.

In a previous study, we introduced a neural network based implementation of this frame-work, and demonstrated its applicability to several spatial navigation and non-spatial linguistic tasks \cite{stoewer_neural_2022}. Here, we further extended our model as shown in Figure \ref{fig1}. In particular, we build a neural network which learns the SR for a non-spatial navigation task based on input feature vectors representing different animal species. The semantic feature vectors represent memory traces, and therefore combine the memory trace theory with the cognitive map theory. 

\section*{Methods}

\subsection*{Successor Representation}
The developed cognitive map is based on the principle of the successor representation (SR). As proposed by Stachenfeld and coworkers the SR can model the firing patterns of the place cells in the hippocampus \cite{stachenfeld2017hippocampus}. The SR was originally designed to build a representation of all possible future rewards $V(s)$ that may be achieved from each state $s$ within the state space over time \cite{SR_Original}. The future reward matrix $V(s)$ can be calculated for every state in the environment, whereas the parameter $t$ indicates the number of time steps in the future that are taken into account, and $R(s_t)$ is the reward for state $s$ at time $t$. The discount factor $\gamma[0,1]$ reduces the relevance of states $s_t$ that are further in the future relative to the respective initial state $s_0$ (cf. eq. \ref{successor_representation}).  

\begin{equation}\label{successor_representation}
\centering
V(s) = E[\sum^{\infty}_{t=0}\gamma^t R(s_t)|s_0=s] 
\end{equation}
Here, $E[\,]$ denotes the expectation value.

The future reward matrix $V(s)$ can be re-factorized using the SR matrix $M$, which can be computed from the state transition probability matrix $T$ of successive states (cf. eq. \ref{SR_refactor}). In case of supervised learning, the environments used for our model operate without specific rewards for each state. For the calculation of these SR we set $R(s_t)=1$ for every state.

\begin{align}\label{SR_refactor}
    V(s) = \sum_{s'} M(s,s')R(s') && M = \sum^{\infty}_{t=0}\gamma^t T^t
\end{align}

\subsection*{Animal Data Set}
The construction of the cognitive map is based on a data set which quantifies seven different semantic features of 32 animal species (Table \ref{trainingdata}). The corresponding test data set is shown in Table \ref{testdata}.

\begin{table}[]
\resizebox{\textwidth}{!}{\begin{tabular}{|l|l|l|l|l|l|l|l|}
\hline
\textbf{Name}   & \textbf{Height(cm)} & \textbf{Weight(kg)} & \textbf{Number Legs} & \textbf{Danger(subjective)} & \textbf{Reproduction (2=Birth, 1= Eggs)} & \textbf{Fur(2=No,1=Yes)} & \textbf{Lungs(2=No,1=Yes)} \\ \hline
Elephant        & 350                 & 6000                & 4                    & 60                          & 2                                        & 2                        & 1                          \\ \hline
Tiger           & 100                 & 100                 & 4                    & 100                         & 2                                        & 1                        & 1                          \\ \hline
Lion            & 120                 & 175                 & 4                    & 100                         & 2                                        & 1                        & 1                          \\ \hline
Dog             & 70                  & 30                  & 4                    & 20                          & 2                                        & 1                        & 1                          \\ \hline
Rabbit          & 40                  & 2                   & 4                    & 0                           & 2                                        & 1                        & 1                          \\ \hline
Bear            & 200                 & 500                 & 4                    & 60                          & 2                                        & 1                        & 1                          \\ \hline
Cow             & 120                 & 500                 & 4                    & 20                          & 2                                        & 1                        & 1                          \\ \hline
Deer            & 70                  & 20                  & 4                    & 0                           & 2                                        & 1                        & 1                          \\ \hline
Cat             & 30                  & 4                   & 4                    & 5                           & 2                                        & 1                        & 1                          \\ \hline
Beaver          & 60                  & 25                  & 4                    & 5                           & 2                                        & 1                        & 1                          \\ \hline
Giraffe         & 500                 & 1200                & 4                    & 40                          & 2                                        & 1                        & 1                          \\ \hline
Ape             & 70                  & 40                  & 4                    & 30                          & 2                                        & 1                        & 1                          \\ \hline
Horse           & 120                 & 250                 & 4                    & 10                          & 2                                        & 1                        & 1                          \\ \hline
Camel           & 125                 & 400                 & 4                    & 10                          & 2                                        & 1                        & 1                          \\ \hline
Goat            & 70                  & 60                  & 4                    & 5                           & 2                                        & 1                        & 1                          \\ \hline
Sheep           & 60                  & 20                  & 4                    & 5                           & 2                                        & 1                        & 1                          \\ \hline
Pig             & 60                  & 200                 & 4                    & 5                           & 2                                        & 2                        & 1                          \\ \hline
Hamster         & 5                   & 0.2                 & 4                    & 0                           & 2                                        & 1                        & 1                          \\ \hline
Dolphine        & 200                 & 60                  & 0                    & 10                          & 2                                        & 2                        & 1                          \\ \hline
Raccoon         & 50                  & 15                  & 4                    & 5                           & 2                                        & 1                        & 1                          \\ \hline
Red Pander      & 30                  & 5                   & 4                    & 5                           & 2                                        & 1                        & 1                          \\ \hline
Ant             & 0,1                 & 0,00001             & 6                    & 1                           & 1                                        & 2                        & 2                          \\ \hline
Bee             & 1                   & 0,0001              & 6                    & 5                           & 1                                        & 2                        & 2                          \\ \hline
Cockroach       & 5                   & 0,005               & 6                    & 0                           & 1                                        & 2                        & 2                          \\ \hline
Goliathus       & 8                   & 5                   & 6                    & 0                           & 1                                        & 2                        & 2                          \\ \hline
Giant weta      & 10                  & 0,035               & 6                    & 0                           & 1                                        & 2                        & 2                          \\ \hline
Heteropteryx    & 15                  & 0,05                & 6                    & 0                           & 1                                        & 2                        & 2                          \\ \hline
Cane toad       & 15                  & 1                   & 4                    & 0                           & 1                                        & 2                        & 1                          \\ \hline
Fire Salamander & 17                  & 0,035               & 4                    & 0                           & 1                                        & 2                        & 1                          \\ \hline
Frog            & 4                   & 0,01                & 4                    & 0                           & 1                                        & 2                        & 1                          \\ \hline
Olm             & 20                  & 0,02                & 4                    & 0                           & 1                                        & 2                        & 1                          \\ \hline
Tree Frog       & 4                   & 0,005               & 4                    & 0                           & 1                                        & 2                        & 1                          \\ \hline
\end{tabular}}
\caption{\textbf{Training data set used to create the cognitive room.} It consists of 32 different animal species, which belong to three different taxonomic classes: mammals, insects and amphibians. Each animal is characterized by seven semantic features: Height, weight, number of legs, its danger level, the reproduction system, if it has fur and if it has lungs.}
\label{trainingdata}
\end{table}

\begin{table}[]
\resizebox{\textwidth}{!}{\begin{tabular}{|l|l|l|l|l|l|l|l|}
\hline
\textbf{Name} & \textbf{Height(cm)} & \textbf{Weight(kg)} & \textbf{Number Legs} & \textbf{Danger(subjective)} & \textbf{Reproduction (2=Birth, 1= Eggs)} & \textbf{Fur(2=No,1=Yes)} & \textbf{Lungs(2=No,1=Yes)} \\ \hline
Jaguar       & 70                  & 70                  & 4                    & 90                          & 2                                        & 1                        & 1                          \\ \hline
Donkey        & 100                 & 200                 & 4                    & 10                          & 2                                        & 1                        & 1                          \\ \hline
Wild boar     & 70                  & 180                 & 4                    & 20                          & 2                                        & 1                        & 1                          \\ \hline
Melontha      & 2,5                 & 0,001               & 6                    & 0                           & 1                                        & 2                        & 2                          \\ \hline
Dragonfly     & 6                   & 0,0003              & 6                    & 0                           & 1                                        & 2                        & 2                          \\ \hline
Wasp          & 1,5                 & 0,00008             & 6                    & 20                          & 1                                        & 2                        & 2                          \\ \hline
\end{tabular}}
\caption{\textbf{Test data used to evaluate the interpolation capabilities of the trained neural network.} It consists of 6 different animal species, which belong to three different taxonomic classes: mammals, insects and amphibians. Again, each animal is characterized by seven semantic features: Height, weight, number of legs, its danger level, the reproduction system, if it has fur and if it has lungs.}
\label{testdata}
\end{table}

The data matrix represents the memory matrix $M(m)$, which our cognitive map is based on. Therefore every animal represents a past memory, and reflects a state in our model. To use the matrix for our supervised learning approach, we need to sample successor labels for every state, which reflect the similarity between animal species. We choose to use the Euclidean distance to calculated the transition probabilities for our state space. Therefore animal species sharing similar semantic features have a higher state transition probability. 

\begin{align}\label{state_transitions}
   T(s,s')=\frac{1}{||m_s-m_{s'}||}
\end{align}

For the generation of the training and test data set, a random starting state is chosen and also a random probability ranging from 0 to 1 is sampled. The input feature vector for the chosen input state is altered by a random range of 0 to 15\,\% to make the training more robust to novel inputs. Based on the sampled probability and the cumulative density function of the defined successor representation matrix, a valid successor state is randomly drawn as label. 10\,\% of the generated samples are not used for training, but are instead preserved as validation data set.

\subsection*{Neural network architectures and training parameters}

\begin{figure}[htbp]
    \centering
    \includegraphics[width = 15cm]{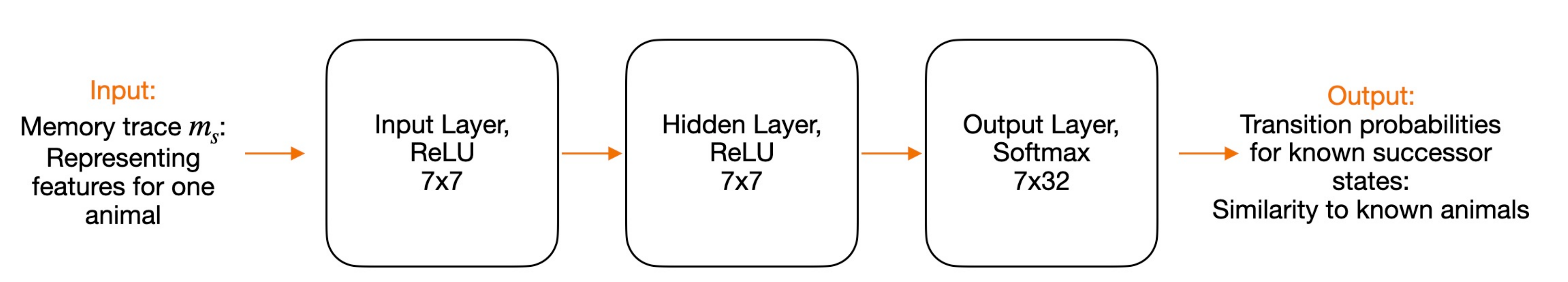}
    \caption{\textbf{Architecture of the trained neural network.} The network receives a memory trace of animal features as input. The size of the input and hidden layer is equal to the number of features in the input. The output layer is a softmax layer with 32 neurons, matching the number of memory traces in the training memory matrix. The output of the network is the probability of the similarity of the input to the entries of the memory matrix used during training.}
    \label{fig2}
\end{figure}

We set up a three-layered feed forward neural network. The network consists of 7 input neurons and has 7 neurons in the hidden layer. Both use a ReLU activation function. The output layer consists of 32 neurons with a softmax activation function (cf. \ref{fig2}). The networks learns the transition probabilities of the environment, i.e. in our case the memory space. Smaller number of neurons in the hidden layer did not influence the results in previous experiments \cite{stoewer_neural_2022}. 
We trained three networks for different discount factors of the successor representation, with $\gamma= (0.3,0.7,1.0)$ and $t=10$. Note that, larger discount factors correspond to a larger time horizon, i.e. taking into account more future steps. 
The networks were trained for 500 epochs, with a batch size of 50, 50000 training samples, using the Adam optimizer with a learning rate of 0.001 and categorical cross-entropy as loss function.

\subsubsection*{Transition probability and successor representation matrix}
After the training process, the networks can predict all probabilities for the successor states for any given input feature vector. Concatenating the predictions of all known animal states leads to the successor representation matrix of the cognitive room. The output of the network is a vector shaped like a row of the respective environment's SR matrix and can therefore directly be used to fill the SR matrix, respectively.

\subsubsection*{Interpolating unknown features}
We propose that the successor representation can be used as a pointer to stored memories. In our case we have the saved memories of 32 animal species in the memory trace matrix which we use for training the network.
If incomplete information is fed into the network (unknown values set to $-1$ in the input feature vector), it still outputs predictions for the possible transition probabilities. 

\begin{align}\label{interpolation}
   m_{interpolated}= SR_{prediction}*M(m_s)
\end{align}

Thus, we can use the prediction from the network, and perform a matrix multiplication with our known memory matrix in order to derive an interpolated feature vector for the incomplete or unknown input (cf. \ref{interpolation}).

\subsection*{Multi-dimensional scaling}
A frequently used method to generate low-dimensional embeddings of high-dimensional data is t-distributed stochastic neighbor embedding (t-SNE) \cite{van2008visualizing}. However, in t-SNE the resulting low-dimensional projections can be highly dependent on the detailed parameter settings \cite{wattenberg2016use}, sensitive to noise, and may not preserve, but rather often scramble the global structure in data \cite{vallejos2019exploring, moon2019visualizing}.
In contrast to that, multi-Dimensional-Scaling (MDS) \cite{torgerson1952multidimensional, kruskal1964nonmetric,kruskal1978multidimensional,cox2008multidimensional} is an efficient embedding technique to visualize high-dimensional point clouds by projecting them onto a 2-dimensional plane. Furthermore, MDS has the decisive advantage that it is parameter-free and all mutual distances of the points are preserved, thereby conserving both the global and local structure of the underlying data. 

When interpreting patterns as points in high-dimensional space and dissimilarities between patterns as distances between corresponding points, MDS is an elegant method to visualize high-dimensional data. By color-coding each projected data point of a data set according to its label, the representation of the data can be visualized as a set of point clusters. For instance, MDS has already been applied to visualize for instance word class distributions of different linguistic corpora \cite{schilling2021analysis}, hidden layer representations (embeddings) of artificial neural networks \cite{schilling2021quantifying,krauss2021analysis}, structure and dynamics of recurrent neural networks \cite{krauss2019analysis, krauss2019recurrence, krauss2019weight}, or brain activity patterns assessed during e.g. pure tone or speech perception \cite{krauss2018statistical,schilling2021analysis}, or even during sleep \cite{krauss2018analysis,traxdorf2019microstructure}. 
In all these cases the apparent compactness and mutual overlap of the point clusters permits a qualitative assessment of how well the different classes separate.

\subsection*{Code Implementation}
The models were coded in Python 3.10. The neural networks were design using the Keras \cite{keras} library with TensorFlow \cite{tensorflow2015-whitepaper}. Mathematical operations were performed with numpy \cite{numpy} and scikit-learn \cite{scikit-learn} libraries.
Visualizations were realised with matplotlib \cite{matplot}.

\section*{Results}

\subsection*{Learning structures by observing states and their successors}

\begin{figure}[htbp]
    \centering
    \includegraphics[width=1.0\linewidth]{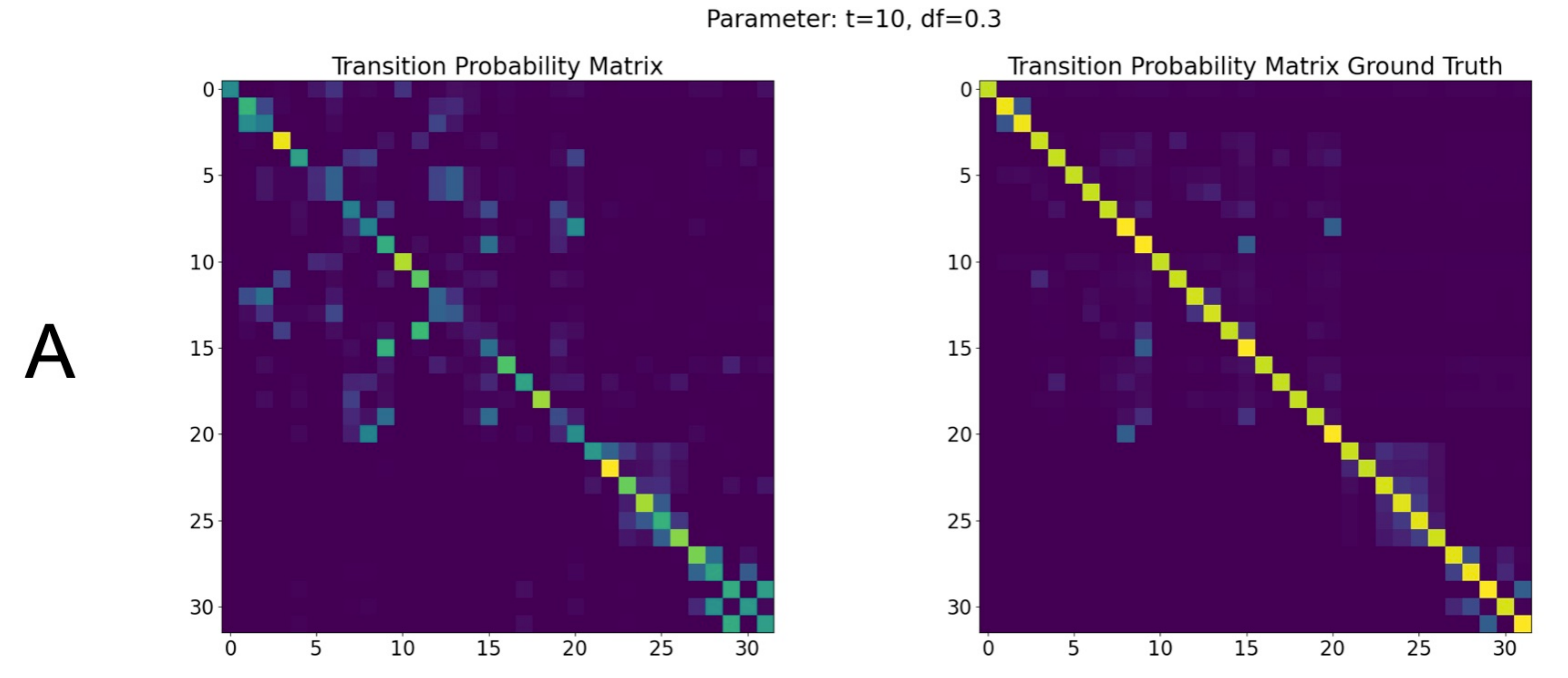}
    \includegraphics[width=1.0\linewidth]{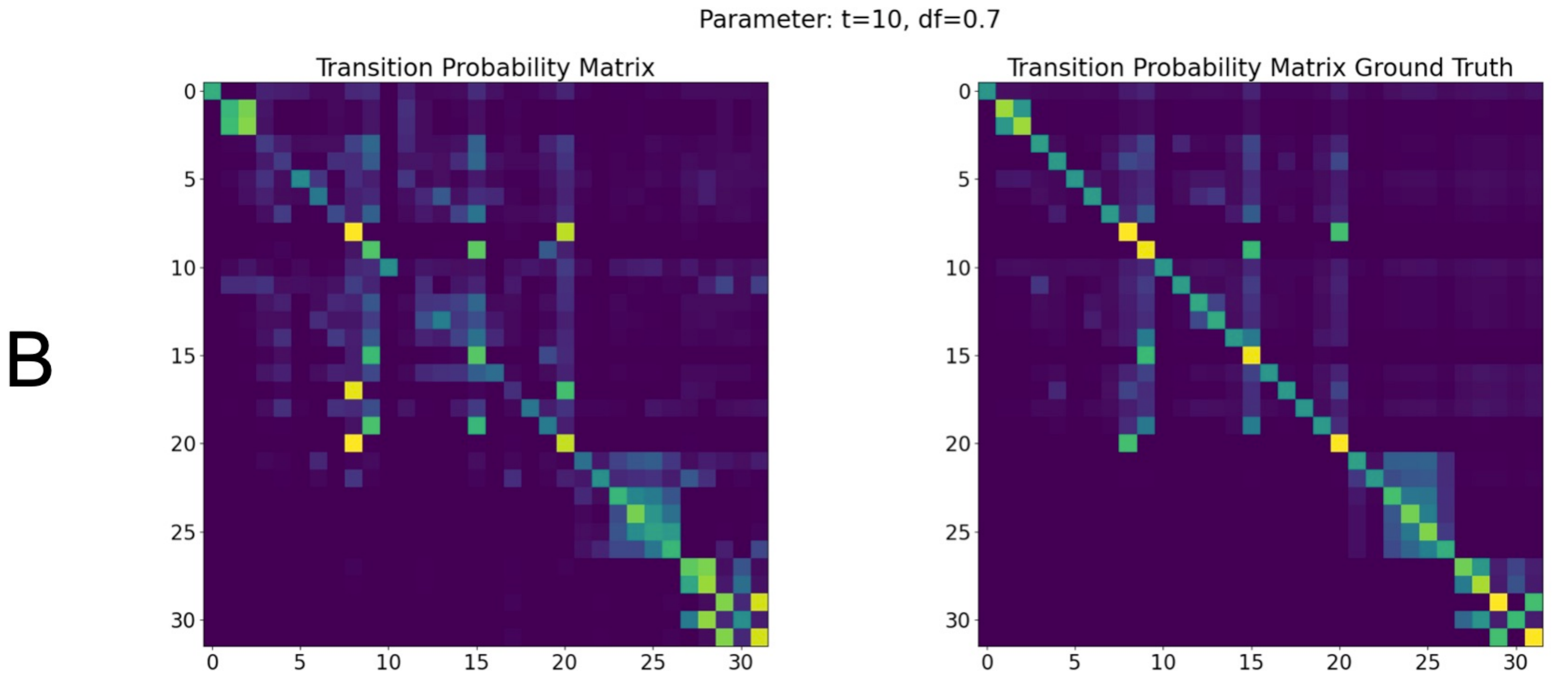}
    \includegraphics[width=1.0\linewidth]{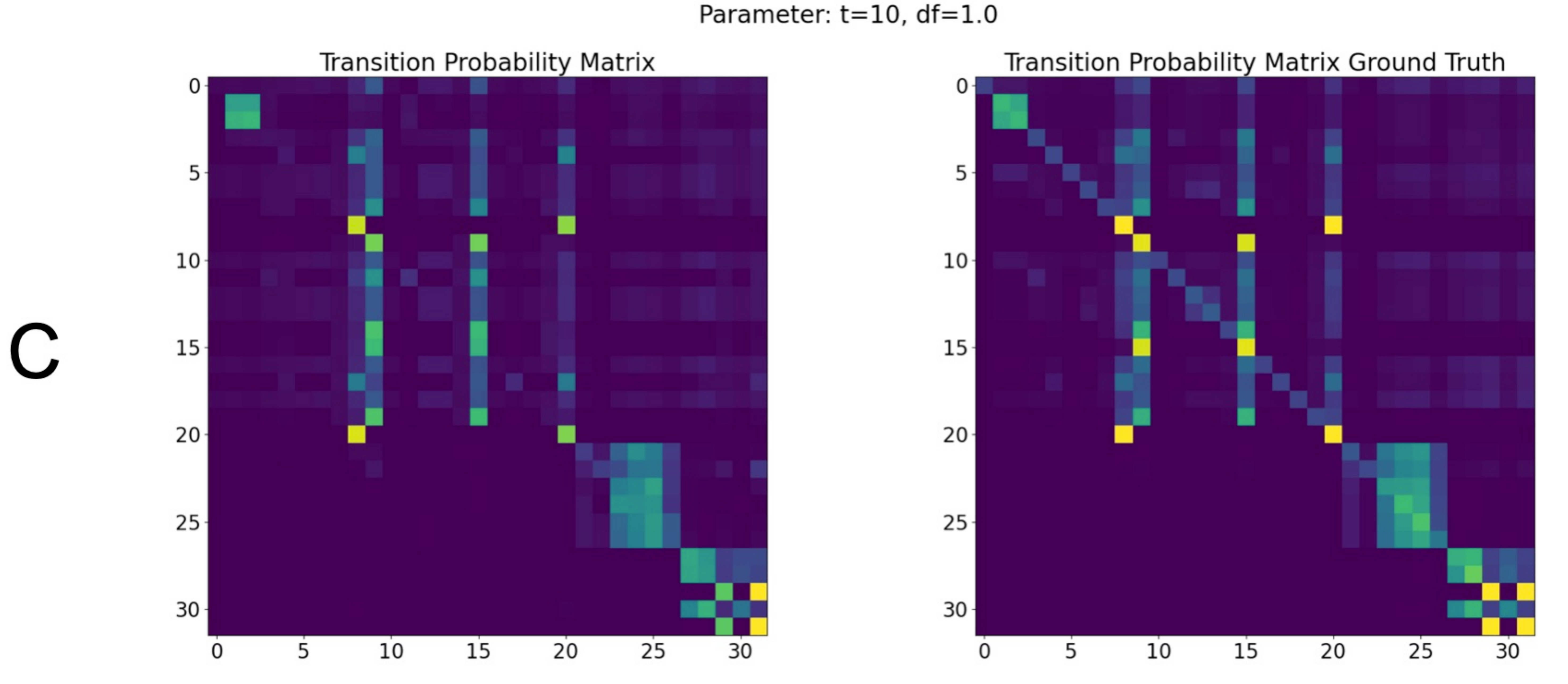}
    \caption{\textbf{Learned successor representation (SR) matrices and corresponding ground truths.} Learned SR matrices (left column) are very similar to their corresponding ground truth SR matrices (right column). a: For a discount factor of $\gamma=0.3$, the RMSE between learned and ground truth SR matrix is $0.02034$. b: For $\gamma=0.7$, the RSME is $0.01496$. c: For $\gamma=1.0$, the RMSE is $0.00854$.}
    \label{fig3}
\end{figure}

The models were trained to learn the underlying structure of the data set. In particular, we trained three different neural networks using different discount factors ($\gamma= (0.3,0.7,1.0)$, $t=10$). The resulting successor representation matrices for each parameter setting are very similar to the ground truth (Figure \ref{fig3}), and the corresponding root-mean-squared errors (RMSE) are extremely low: $0.02034$ for $\gamma=0.3$ (Figure \ref{fig3}a), $0.01496$ for $\gamma/0.7$ (Figure \ref{fig3}b), and $0.00854$ for $\gamma=1.0$ (Figure \ref{fig3}c). 

\begin{figure}
\centering
\includegraphics[width=1.0\linewidth]{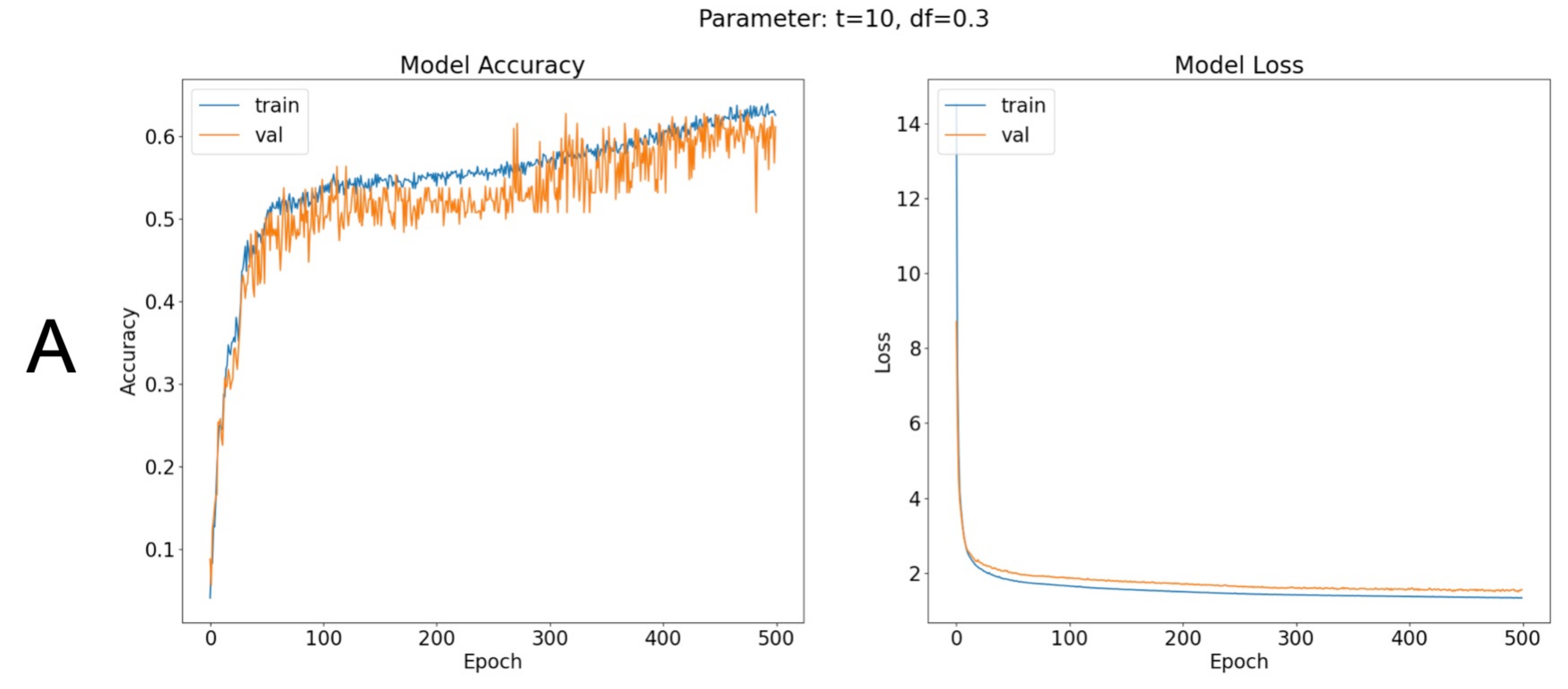}\\
\includegraphics[width=1.0\linewidth]{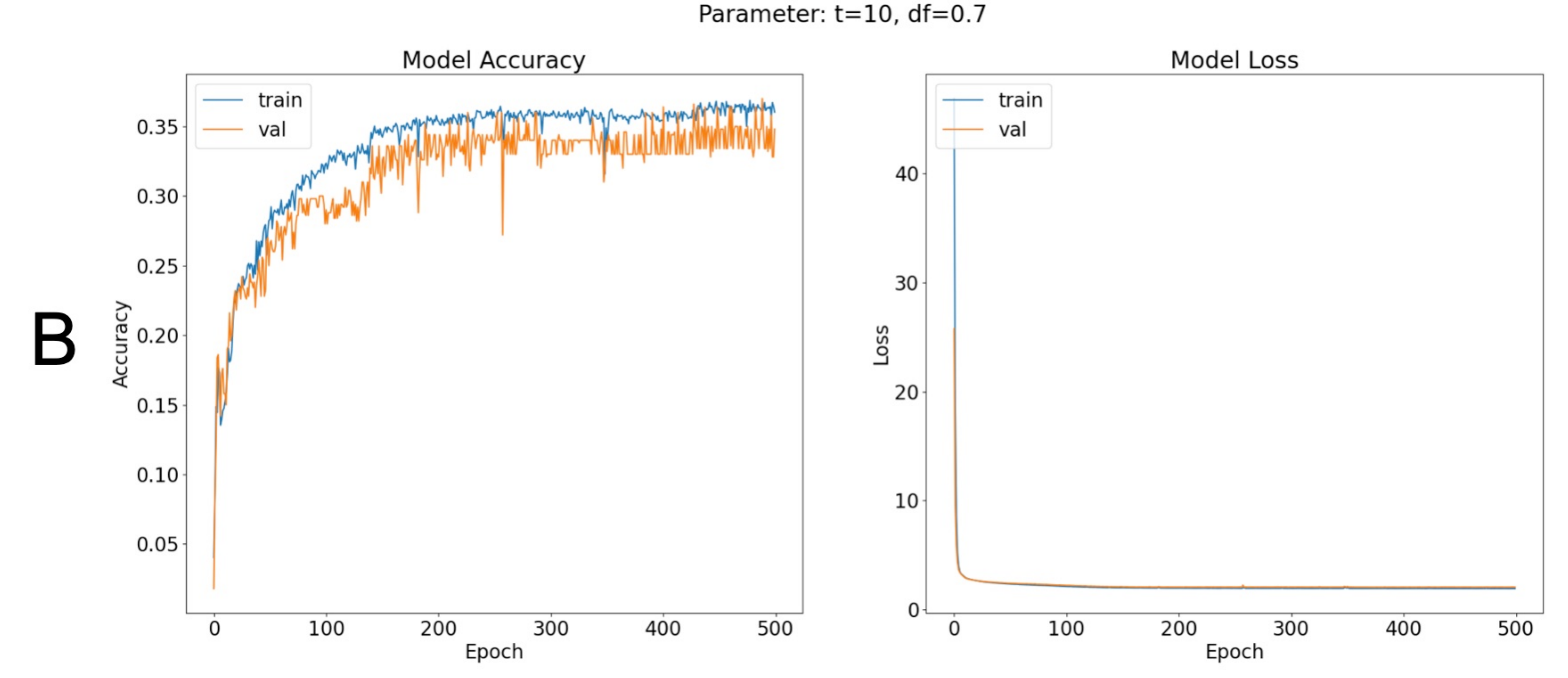}\\
\includegraphics[width=1.0\linewidth]{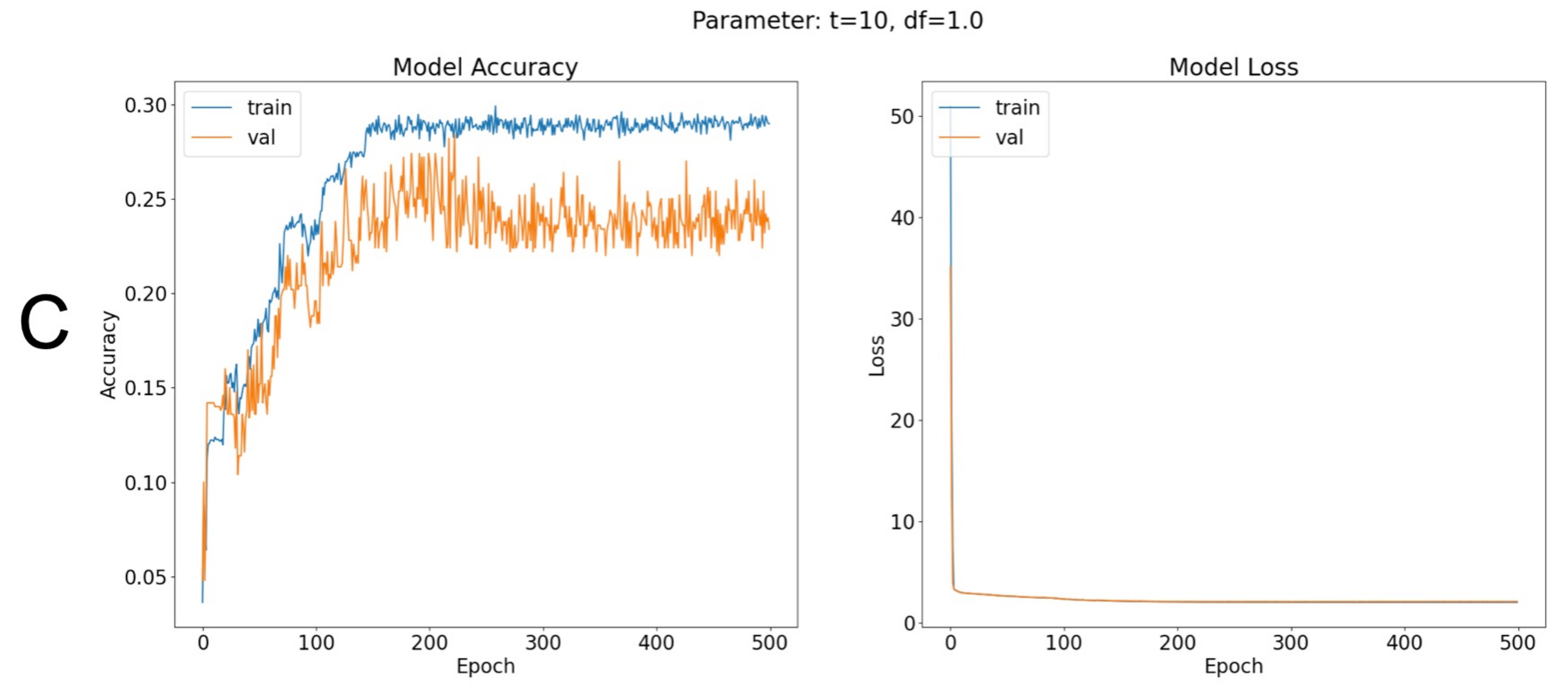}
\caption{\textbf{Accuracies and loss for different models.} Training accuracies (blue) and validation accuracies (orange) during training are shown in the left column. The corresponding loss is shown in the right column. a: For a discount factor of $\gamma = 0.3$, the highest accuracy of $60\%$ was achieved. b: For $\gamma=0.7$, the accuracy saturates after 200 epochs at $30\%$. c: For $\gamma=1.0$, the accuracy saturates after 200 epochs at $35\%$.}
\label{fig4}
\end{figure}

The accuracy for the model with the discount factor $\gamma=0.3$ increased quickly during the first 100 epochs, and then slowly continued to increase until the end of training at epoch 500 where the highest accuracy of $60\%$ was achieved (Figure \ref{fig4}a). 

In contrast, the training procedure quickly reached a saturation of the accuracy for the two models with discount factors $\gamma=0.7$ and $\gamma=1.0$ after around 200 epochs, with maximum training and validation accuracies of approximately $30\%$ or $35\%$ respectively (Figure \ref{fig4}b, c).

\subsection*{Scaling of cognitive maps depends on discount factor of successor representation}

The discount factor of the SR is proposed to enable scaling of the cognitive maps, and thus to represent hierarchical structures, similar to the different mesh sizes of grid cells along the longitudinal axis of the hippocampus and the entorhinal cortex \cite{stachenfeld2017hippocampus}. Actually, memory representations, such as the internal representation of space, systematically vary in scale along the hippocampal long axis \cite{collin2015memory}. This scaling has been suggested to be used for targeted navigation with different horizons \cite{brunec2019predictive} or even for encoding information from smaller episodes or single objects to more complex concepts \cite{milivojevic2013mnemonic}.

\begin{figure}[htbp]
    \centering
    \includegraphics[width = 17cm]{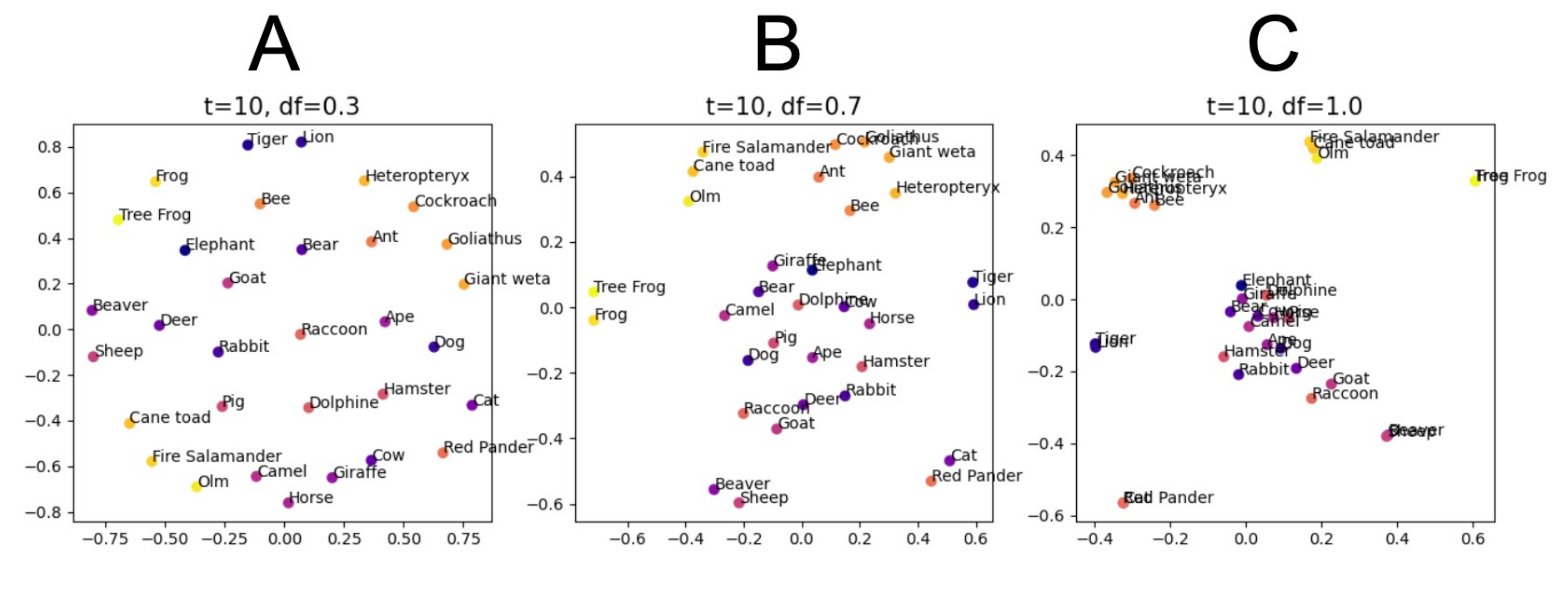}
    \caption{\textbf{Different scalings of cognitive maps.} Shown are MDS projections of SR matrices with different discount factors $\gamma$. a: For a low discount factor $\gamma=0.3$, the resulting map is most fine-grained and detailed with little clustering (GDV: $-0.322$). b: A medium discount factor of $\gamma=0.7$ results in an intermediate scale with more clustering (GDV: $-0.355$ ) compared to (a). c: The largest discount factor $\gamma=1.0$ results in the most coarse-grained map. Here, individual animal species are no longer distinguishable, but instead form different clusters corresponding to more abstract concepts, i.e. the taxonomic animal classes mammals (blue, purple), insects (orange), and amphibians (yellow) (GDV: $-0.403$). Note that, a GDV of $-1.0$ indicates perfect clustering, whereas a GDV of $0.0$ indicates no clustering at all.}
    \label{fig5}
\end{figure}

In order to visualize the learned SR underlying the cognitive maps, we calculate MDS pojections from the SR matrices (Figure \ref{fig5}). Furthermore, as an estimate for the map scaling, we calculate the general discrimination value (GDV, cf. Methods) for each map. 

We find that the resulting scaling of the cognitive maps depends on the discount factor of the underlying SR matrix, and that the GDV correlates with the discount factor. A small discount factor of $\gamma=0.3$ results in a fine-grained and detailed cognitive map where each object is clearly separated from the others, and similar objects, i.e. animal species, are closer together (Figure \ref{fig5}a). With a GDV of $-0.322$, the clustering is relatively low compared to the other maps. This cognitive map resembles so called self-organizing maps introduced by Kohonen \cite{kohonen1990self}, and might correspond to word fields proposed in linguistics \cite{aitchison2012words}.

A discount factor $\gamma=0.7$ results in an intermediate scale cognitive map with a GDV of $-0.355$ (Figure \ref{fig5}b). 

Finally, a discount factor of $\gamma=1.0$ results in the most course-grained cognitive map. Here, individual animal species are no longer clearly separated from each other, but are forming instead representational clusters that correspond to taxonomic animal classes, i.e. mammals, insects and amphibians (Figure \ref{fig5}c). Consequently, these map has the lowest GDV of $-0.403$, indicating the best clustering. This type of representation generalizing from individual objects might correspond to the emergence of cognitive categories, as suggested e.g. in prototype semantics \cite{cruse2014prototype}.

\subsection*{Feature inference for incomplete feature vectors}

\begin{figure}[htbp]
    \centering
    \includegraphics[width = 15cm]{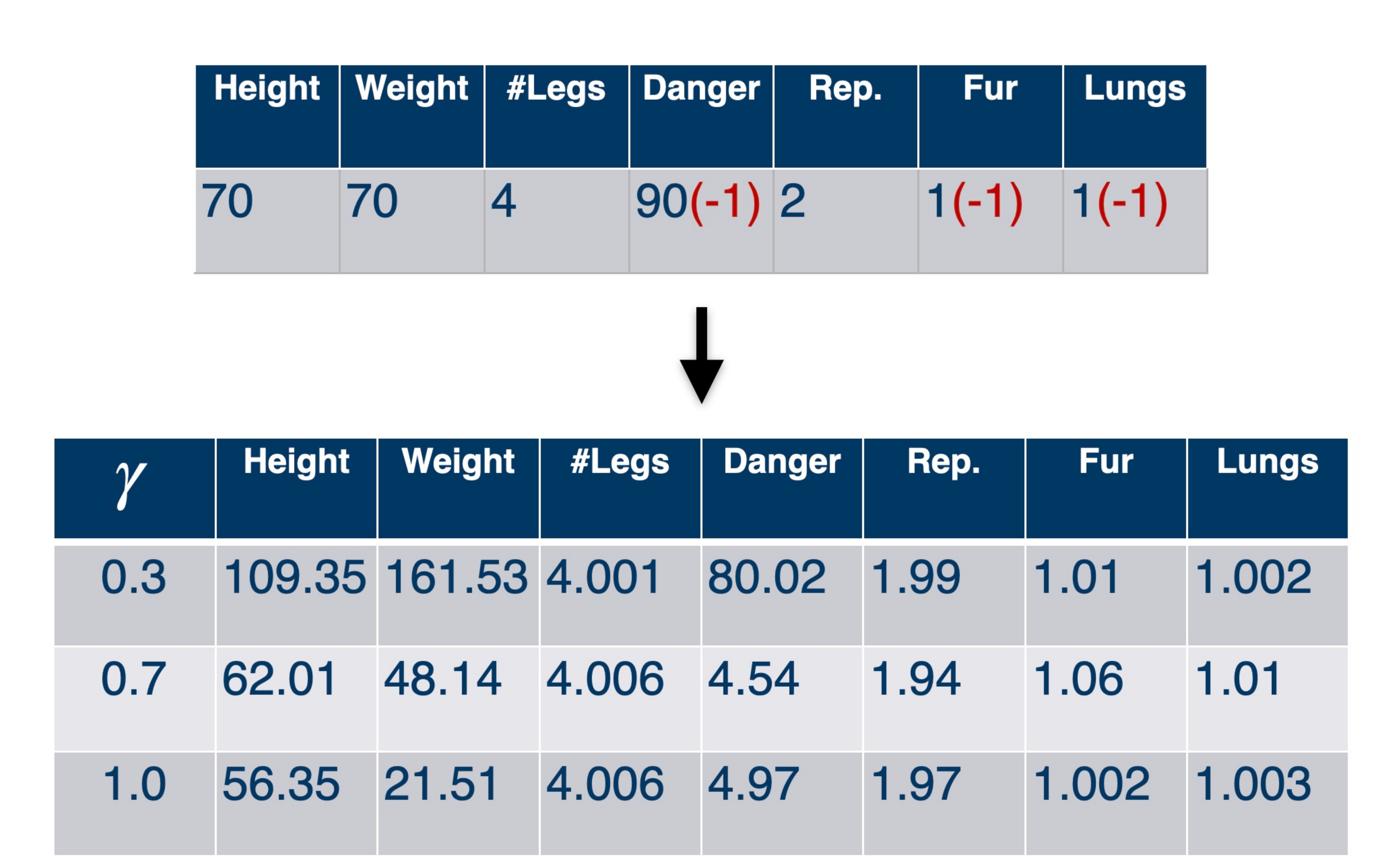}
    \caption{\textbf{Interpolation of the test data set feature vector 'Jaguar'}. Three semantic features (dangerous, having a fur and having lungs) are missing, i.e. are replaced by the value $-1$. The three networks trained with different discount factors  $\gamma= (0.3,0.7,1.0)$ infer the missing features. Binary semantic features are inferred well in all cases. The 'dangerous' feature is badly predicted for large discount factors $\gamma= (0.7,1.0)$. In contrast, in case of the lower discount factor $\gamma= 0.3$, it is predicted well.}
    \label{fig6}
\end{figure}

The neural network which learned the structure of the input data successfully can now be used to interact with unseen data. The prediction of the trained neural network can be used as weighted pointer to the memorized objects (animal species) in the training data set. The vector of a previous unseen animal, the \emph{jaguar}, is feed into the network for prediction \ref{fig6}. Three features (danger, fur, lungs) are missing, i.e. are set to $-1$. The binary features are predicted well independent from the discount factor. The 'danger' feature is inferred best for the smallest discount factor $\gamma = 0.3$. Note that, also the not missing parameters are changed by the prediction. In general, larger discount factors better infer more general features, whereas smaller discount factors better infer more specific features.

\begin{figure}[htbp]
    \centering
    \includegraphics[width = 19cm]{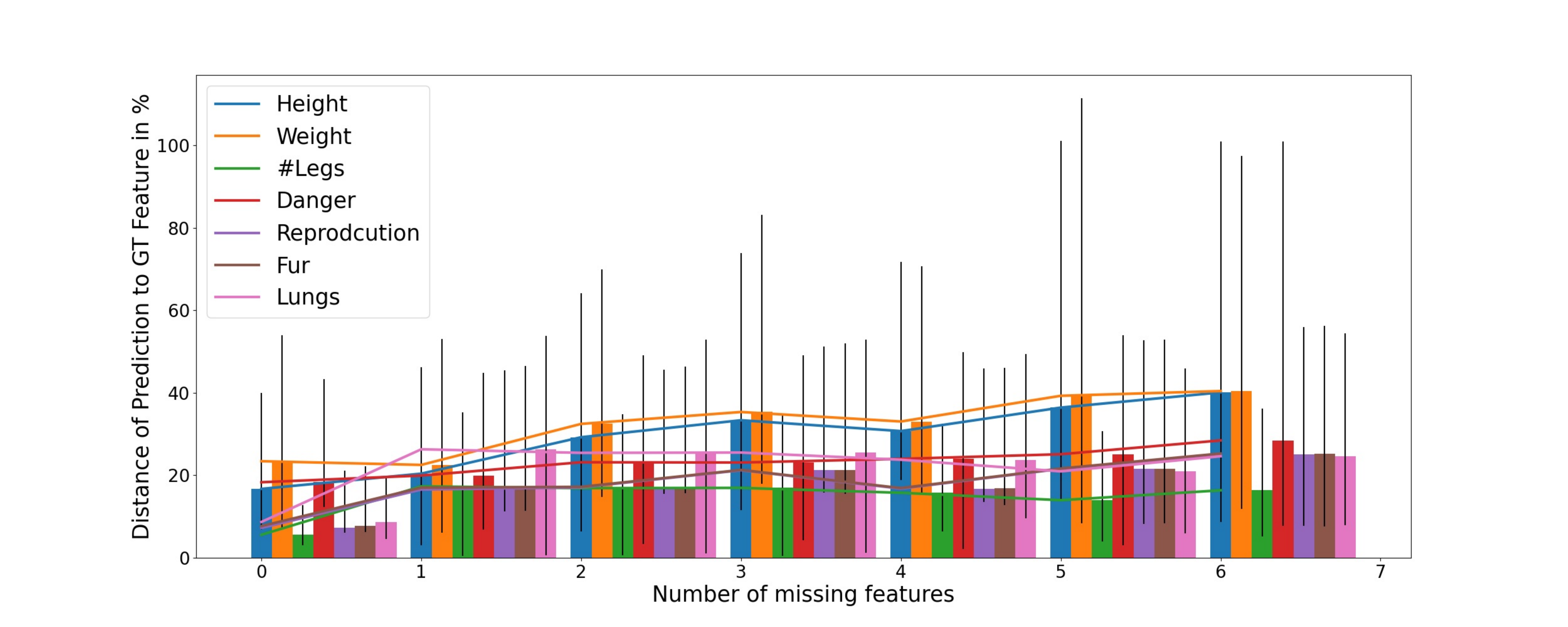}
    \caption{\textbf{Dissimilarities between interpolated features and ground truth.} 10 networks with $\gamma=1.0$ have been trained. Dissimilarity is low in case of a no or a single missing feature, and increases with number of missing features up to 40\% for six missing features. In general, binary semantic features are inferred with better accuracy than non-binary semantic features. The variance of the different networks for the features 'height' and 'weight' are highest.}
    \label{fig7}
\end{figure}

We further evaluated the model with our interpolation test data set (cf. \ref{testdata}). We trained ten models with the parameters $\gamma=1.0$ and $t=10$. In Figure \ref{fig7} the distances of the predictions of different features in comparison to the ground truth is summarized. The percentage is based on the maximum distance of the according feature. The evaluation is plotted for the feature vectors, with up to 6 missing entries for every prediction. The distance of the prediction to ground truth with no missing entries is in general low ranging from around 5\% to 25\% (corresponding to 95\% to 75\% accuracy), indicating high similarity. However, dissimilarity increases to 40\% in case of 6 missing features. The distance is however different for each feature. While the semantic feature 'number of legs' is predicted well, the height of the animal is predicted with less accuracy. Furthermore, the variance differs for different models. Especially the badly predicted predicted features like 'height' and 'weight' the variance is quite large. 

\section*{Discussion}

In this study we have demonstrated that arbitrary feature spaces can be learned efficiently on the basis of successor representations with neural networks. In particular, the networks learn a representation of the semantic feature space of animal species as a cognitive map. The network achieves an accuracy of around 30\% which is near to the theoretical maximum regarding the fact that all animal species have more than one possible successor, i.e. nearest neighbor in feature space. Our approach therefore combines the concepts of feature space and state space. The emerging representations therefore resemble the proposed general and abstract cognitive maps described by Bellmund et al. \cite{bellmund2018navigating}. 

Our model extends our past work, where we reproduced place cell fire patterns in a spatial navigation and a non-spatial linguistic task based on the successor representation \cite{stoewer_neural_2022}. The innovation of our here presented approach is, that we can use the successor representation with arbitrary new input as a weighted pointer to already stored input data, i.e. memories, thereby combining the two hallmarks of hippocampal processing: declarative memory and navigation. The successor representation might therefore be a tool which can be used to navigate through arbitrary cognitive maps, and find similarities in novel inputs as well as past memories. 

Furthermore, the discount factor $\gamma$ of the successor representation can be used to model cognitive maps with different scales, which range in our example from clusters of taxonomic animal classes to individual animal species. The varying grid cell scaling along the long axis of the entorhinal cortex is known to be associated with hierarchical memory content  \cite{collin2015memory}. The discount factor can be used to model this hierarchical structure. In our experiment the hierarchical scale could be used to interpolate novel feature data in different ways. For example if we want to retrieve general information, a large discount factor resulting in dense clusters, to derive averaged information about the whole cluster, can be used. In contrast, for more detailed information regarding a specific state of the cognitive map, a smaller discount factor is useful. 

Since our approach works with a direct feature vector as input, it still requires highly pre-processed data. A future outlook for this model could be to include a deep neural network for feature extraction as pre-processing. For instance, image analysis is already a well established field for deep neural networks. Our model could be used to replace the last output layer of such networks, which usually perform a classification task, and use the feature space embeddings to learn a cognitive map. This extended model could enhance the learning from just classification to understanding which features are present in which image. This could potential lead to more context awareness in neural networks.

As recently suggested, the neuroscience of spatial navigation might be of particular importance for artificial intelligence research \cite{bermudez2020neuroscience}. A neural network implementation of hippocampal successor representations, especially, promises advances in both fields. Following the research agenda of Cognitive Computational Neuroscience proposed by Kriegeskorte et al. \cite{kriegeskorte2018cognitive}, neuroscience and cognitive science benefit from such models by gaining deeper understanding of brain computations \cite{schilling2020intrinsic, krauss2018cross, krauss2021analysis}. Conversely, for artificial intelligence and machine learning, neural network-based multi-scale successor representations to learn and process structural knowledge as an example of neuroscience-inspired artificial intelligence \cite{hassabis2017neuroscience,krauss2020will,yang2021neural,maier2022known}, might be a further step to overcome the limitations of contemporary deep learning \cite{krauss2012parameter, marcus2018deep, yang2021neural, gerum2021integration, maier2019learning,maier2022known} and towards human-level artificial general intelligence.

\FloatBarrier
\section*{Acknowledgments}

This work was funded by the Deutsche Forschungsgemeinschaft (DFG, German Research Foundation): grant KR\,5148/2-1 to PK (project number 436456810) and grant SCHI\,1482/3-1 (project number 451810794) to AS.

\section*{Author contributions}
PS performed computer simulations and prepared all figures. PS, AM and PK designed the study. PK and AM supervised the study. PS, AS, AM and PK discussed the results and wrote the manuscript.

\section*{Competing interests}
The authors declare no competing financial interests.

\FloatBarrier

\end{document}